\documentclass[,final]{aipproc}
\layoutstyle{6x9}

\newcommand{\PRL}[3]{Phys. Rev. Lett. {\bf {#1}}, {#2} ({#3})}
\newcommand{\PR}[3]{Phys. Rev. {\bf {#1}}, {#2} ({#3})}
\newcommand{\PL}[3]{Phys. Lett. {\bf {#1}}, {#2} ({#3})}

\begin{document}

\title{Acceptance effects in the hyperons global polarization measurement}


\author{Ilya Selyuzhenkov for the STAR Collaboration}{
  address={Physics Department, Wayne State University, 666 W Hancock, Detroit MI 48201, USA}
}

\begin{abstract}
The possible sources of systematic uncertainties in the hyperons global polarization measurement are discussed.
The equation with detector acceptance effects taken into account is provided.
Contribution of the hyperons directed flow into the hyperons global polarization measurement is shown.
The systematic uncertainties of the $\Lambda$ hyperons global polarization
measurement in Au+Au collisions with the STAR detector at RHIC are calculated.
\end{abstract}

\maketitle

\paragraph{Introduction}
One of the most interesting and important phenomena predicted
to occur in non-central relativistic nucleus-nucleus collisions is global system
polarization~\cite{LiangPRL94, Voloshin0410089, Liang0411101}.
This global polarization can have many observable consequences.
It should in particular, lead to global polarization of hyperons,
most easily measurable via their weak decays.
Preliminary results of the $\Lambda$ hyperons global polarization measurement
with the STAR detector at RHIC were recently presented \cite{Selyuzhenkov:2005xa,Selyuzhenkov:2006fc}.
In \cite{Selyuzhenkov:2005xa,Selyuzhenkov:2006fc} the $\Lambda$ hyperons
is reconstructed via decay channel $\Lambda \to p \pi^- $.
The global polarization of hyperons is determined from the angular
distribution of hyperon decay products relative to the system orbital
momentum {\boldmath $L$}:
\begin{eqnarray}
\label{GlobalPolarizationDefinition}
\frac{dN}{d \cos \theta^*} \sim 1+\alpha_H~P_H~\cos \theta^*,
\end{eqnarray}
where $P_H$ is the hyperon global polarization,
$\alpha_H$ is the hyperon decay parameter,
$\theta^*$ is the angle between the system orbital momentum {\boldmath $L$}
and the 3-momentum of the baryon from the hyperon decay 
in the hyperon rest frame.
Since the system angular momentum {\boldmath $L$}  is perpendicular to the reaction plane,
the global polarization can be measured via correlations of the
azimuthal angle $\phi^*_p$ of the hyperon decay product (in the hyperon rest
frame) with respect to the reaction plane angle $\Psi_{RP}$.
Based on the definition (\ref{GlobalPolarizationDefinition}) 
and by using the geometrical relation between the angles, namely
$\cos \theta^* = \sin \theta^*_p \cdot \sin \left( \phi^*_p - \Psi_{RP}\right)$
($\theta^*_p$ is the angle between the hyperon's decay product 3-momentum
in the hyperon's rest frame and the beam direction),
one finds the hyperon polarization is given by (see \cite{Selyuzhenkov:2005xa,Selyuzhenkov:2006fc}):
\begin{eqnarray}
\label{GlobalPolarizationObservable}
P_{H}~=~\frac{8}{\pi\alpha_H}\langle \sin \left( \phi^*_p 
- \Psi_{RP}\right)\rangle.
\end{eqnarray}
The angle brackets in this equation denote averaging over all possible directions
of the hyperon's decay product 3-momentum in its rest frame and over all directions
of the system orbital momentum {\boldmath $L$} or,
in other words, over all possible orientations of the reaction plane.
There are two main sources of systematic uncertainties in this measurement.
The first stems from acceptance effects due to hyperon's reconstruction procedure.
The second is due to uncertainty in the reaction plane determination.
The latter effects are the same as in anisotropic flow measurement \cite{Poskanzer:1998yz}
and proven techniques exist to take these effects into account.

In this paper, we concentrate on acceptance effects
which originates from the hyperons reconstruction procedure.
We derive the equation for acceptance correction function and show
the possible admixture of the hyperons directed flow into global polarization measurement.
The contribution to the global polarization measurement from both effects
are estimated for the case of $\Lambda$ hyperons reconstructed from Au+Au collisions
with the STAR detector at RHIC.

\paragraph{\label{AcceptanceCorrections}Acceptance corrections}

\begin{figure}[h]
\includegraphics[height=.23\textheight]{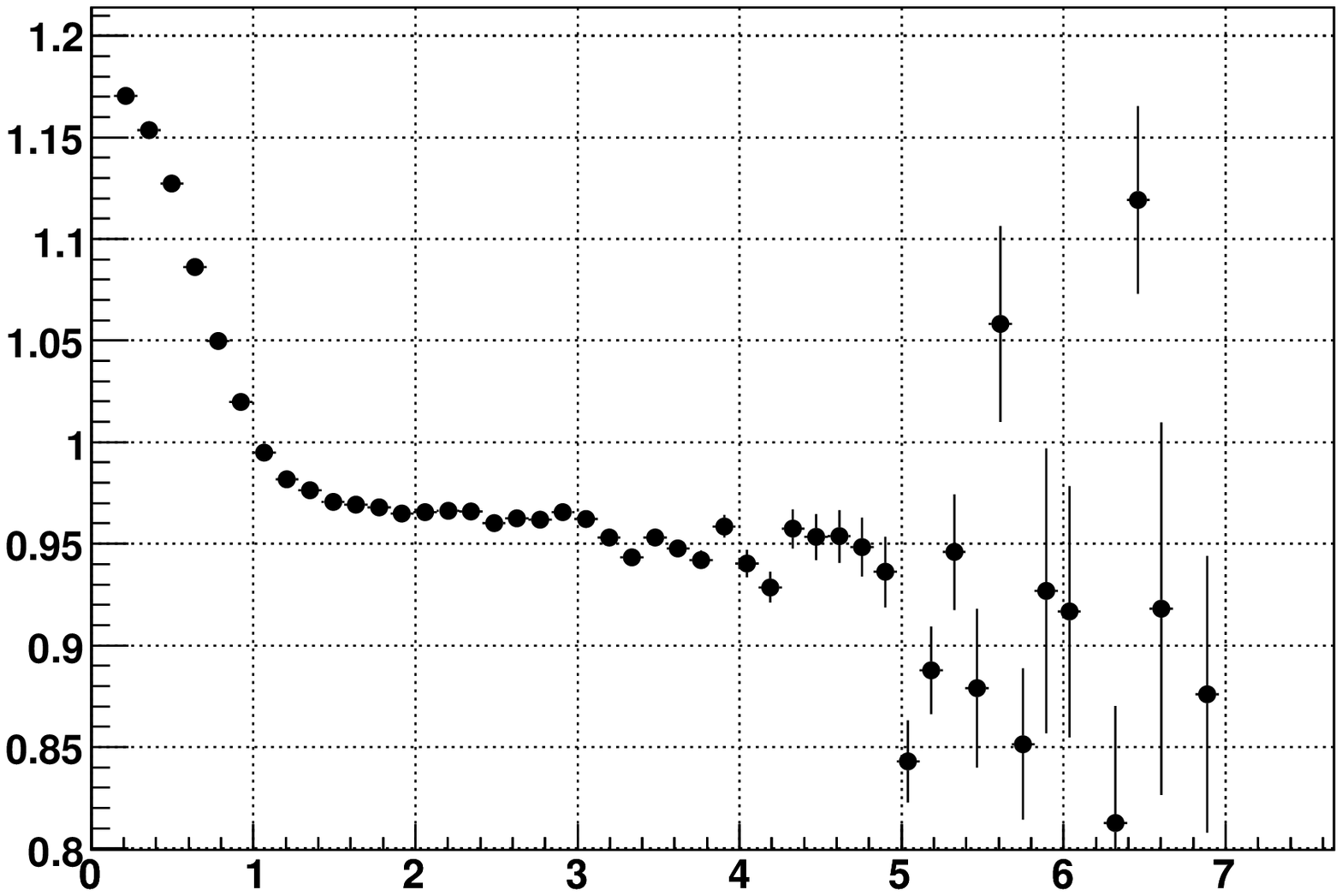}%
\includegraphics[height=.23\textheight]{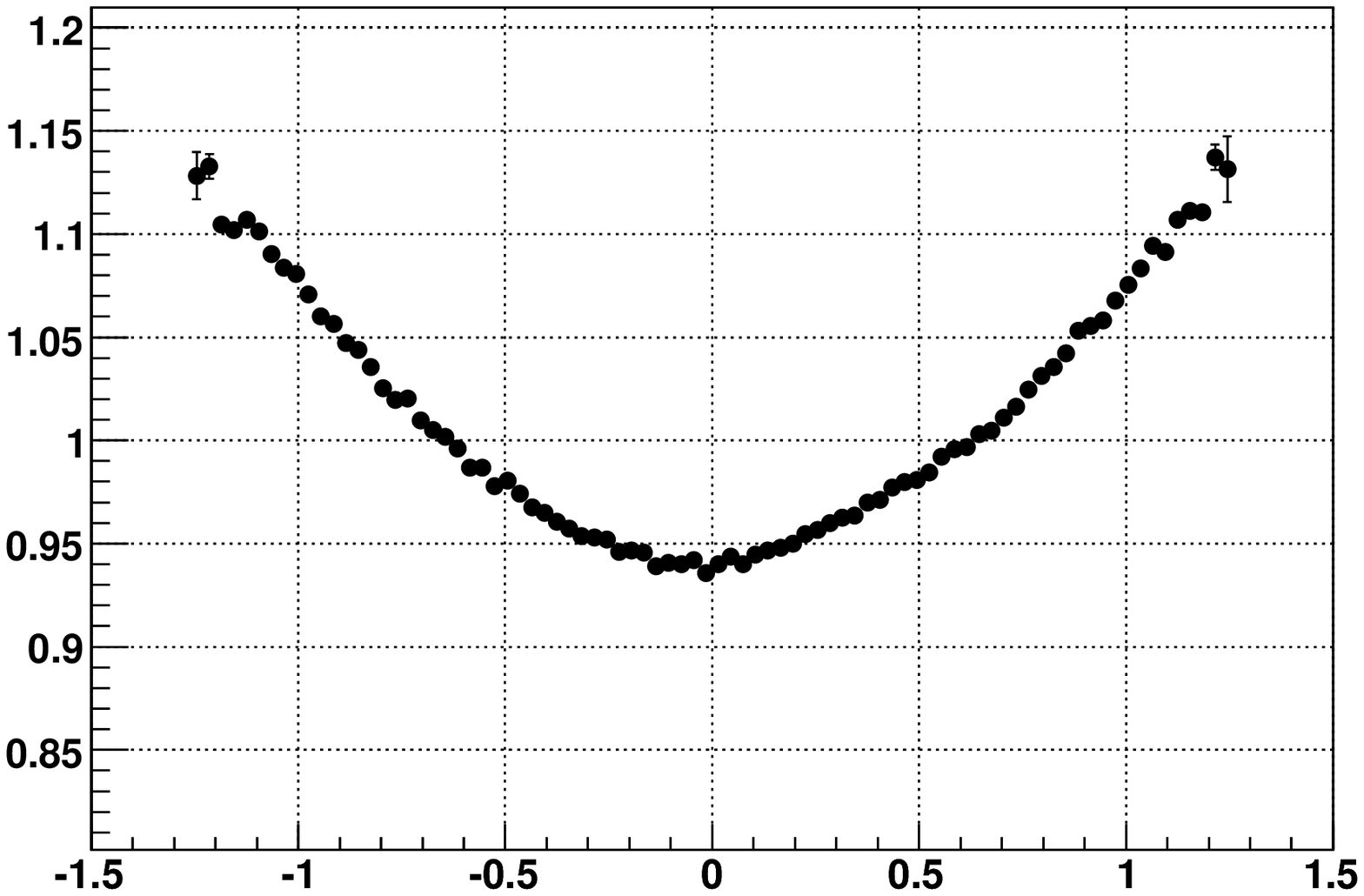}
\put(-220,62){\rotatebox{90}{$A_{corr}^{\Lambda}$}}
\put(-430,62){\rotatebox{90}{$A_{corr}^{\Lambda}$}}
\put(-111,-5){$\eta^{\Lambda}$}
\put(-331,-5){$p_t^{\Lambda}$}
\put(-378,101){STAR Preliminary}
\put(-157,101){STAR Preliminary}
\label{lambdaAccCorrFigure}
\caption{Integral (\ref{AccCoefficient}) as a 
function of $\Lambda$ transverse momentum (left) and pseudo-rapidity (right). See text for details.}
\end{figure}

The derivation of  equation (\ref{GlobalPolarizationObservable})
assumes a perfect reconstruction acceptance for hyperons.
For the case of a non-perfect detector, we similarly consider
the average of $\langle \sin \left( \phi^*_p - \Psi_{RP}\right)\rangle$
but take into account the fact that the integral
over all directions of hyperon's decay products 3-momentum ${\bf p}^*_p$
in the hyperon rest frame is influenced by detector acceptance:
\begin{eqnarray}
\label{meanSinAcc}
\lefteqn{\langle \sin \left( \phi^*_p - \Psi_{RP}\right)\rangle =} \\
\nonumber
&&\int {\frac{d\Omega^*_p}{4\pi}\frac{d\phi_H}{2\pi}
 A({\bf p}_H, {\bf p}^*_p) \int\limits_0^{2\pi} \frac{d\Psi_{RP}}{2\pi} \sin \left( \phi^*_p - \Psi_{RP}\right)}
\left[1+\alpha_H~P_H ({\bf p}_H)~ \sin \theta^*_p \cdot \sin \left( \phi^*_p - \Psi_{RP}\right)\right].
\end{eqnarray}
Here ${\bf p}_H$ is the hyperon 3-momentum,
$d\Omega^*_p = d\phi^*_p \sin \theta^*_p d \theta^*_p$ and
$A\left({\bf p}_H, {\bf p}^*_p\right)$ is a function associated with detector acceptance.
Integration of equation (\ref{meanSinAcc}) over reaction plane angle $\Psi_{RP}$ gives:
\begin{eqnarray}
\label{GlobalPolarizationObservableAcc}
\langle \sin \left( \phi^*_p - \Psi_{RP}\right)\rangle = 
\frac{\alpha_H}{2} \int {\frac{d\Omega^*_p}{4\pi} \frac{d\phi_H}{2\pi}A\left({\bf p}_H, {\bf p}^*_p\right) \sin\theta^*_p P_H({\bf p}_H)}.
\end{eqnarray}
Assuming that global polarization does not depend on hyperon azimuthal direction,
one gets the following expression for the acceptance correction function:
\begin{eqnarray}
\label{AccCoefficient}
A_{corr}(p_t^H,\eta^H) = \frac{4}{\pi} \int {\frac{d\Omega^*_p}{4\pi} \frac{d\phi_H}{2\pi}A\left({\bf p}_H, {\bf p}^*_p\right) \sin\theta^*_p}.
\end{eqnarray}
Note that in case of a detector with perfect acceptance, $A_{corr}(p_t^H,\eta^H) = 1$.
Fig.\ref{lambdaAccCorrFigure} shows the integral (\ref{AccCoefficient})
as a function of pseudo-rapidity $\eta^\Lambda$ and transverse momentum $p_t^\Lambda$ of
$\Lambda$ hyperons reconstructed from Au+Au collisions at $\sqrt{s_{NN}}$=200 GeV with the STAR detector at RHIC.
The deviation of this function from unity is small.
The corresponding corrections to the absolute value of the global polarization are thus estimated to be less than 20\%
and they should not affect the conclusions presented in \cite{Selyuzhenkov:2005xa,Selyuzhenkov:2006fc}.

\begin{figure}[h]
\includegraphics[height=.23\textheight]{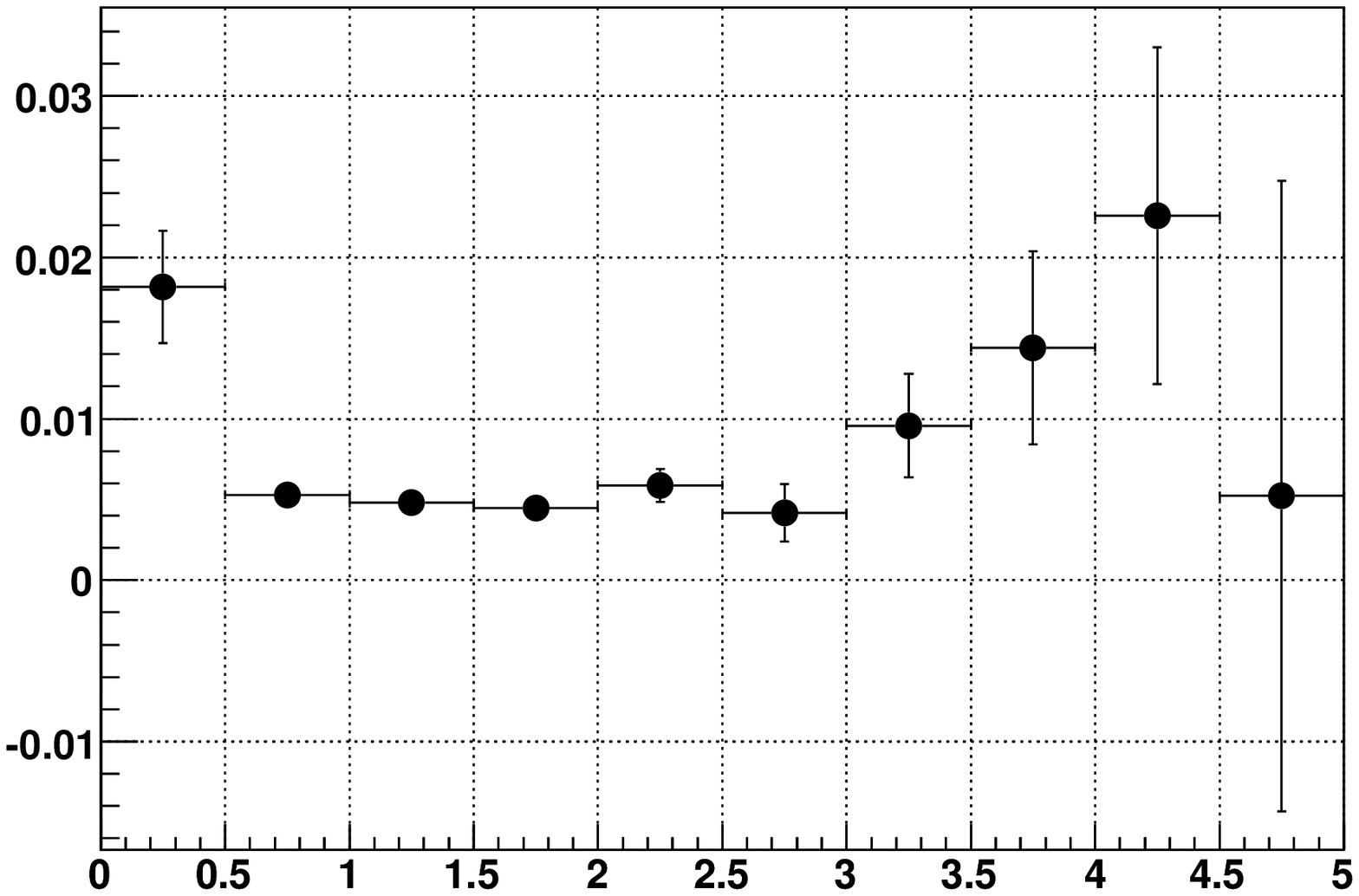}
\put(-220,62){\rotatebox{90}{$A_{corr}^{flow,\Lambda}$}}
\put(-105,-5){$p_t^{\Lambda}$}
\put(-170,103){STAR Preliminary}
\label{v1asymmetry}
\caption{$\Lambda$ hyperon asymmetry $A^{flow, \Lambda}_{asym}$ as a function of $\Lambda$ transverse momentum. See text for details.}
\end{figure}

\paragraph{\label{DirectedFlow}Directed flow admixture}
We derive the hyperons anisotropic flow contribution
to the global polarization measurement
starting with an angular distribution defined by anisotropic flow \cite{Poskanzer:1998yz}:
\begin{eqnarray}
\label{hyperonHlow}
\frac{dN}{d\phi_H} \sim
1+2 \sum_{n=1}^{\infty}~ v_n^H ({\bf p}_H)\cos \left( \phi_H - \Psi_{RP}\right),
\end{eqnarray}
where $v_n^H({\bf p}_H)$ is the hyperon's $n$-harmonic anisotropic flow.
Similarly to Eq.(\ref{meanSinAcc}), we write the expression
for the average of  $\langle \sin \left( \phi^*_p - \Psi_{RP}\right)\rangle$:
\begin{eqnarray}
\label{meanSinVallFlow}
\lefteqn{\langle \sin \left( \phi^*_p - \Psi_{RP}\right)\rangle_{flow} =}\\
\nonumber
&&\int {\frac{d\Omega^*_p}{4\pi} \frac{d\phi_H}{2\pi}
A({\bf p}_H, {\bf p}^*_p) \int\limits_0^{2\pi} \frac{d\Psi_{RP}}{2\pi} \sin \left( \phi^*_p - \Psi_{RP}\right)}
\left[1+2 \sum_{n=1}^{\infty}~ v_n^H ({\bf p}_H)\cos \left( \phi_H - \Psi_{RP}\right)\right],
\end{eqnarray}
where we use the subscript {\it flow} to indicate the different origin of this contribution
to the observable (\ref{GlobalPolarizationObservable}).
Integration of equation (\ref{meanSinVallFlow}) over the reaction plane angle $\Psi_{RP}$ gives
(this integral cuts only terms proportional to first harmonic, which is defined by hyperons directed flow $v_1^H$):
\begin{eqnarray}
\label{meanSinV1}
\langle \sin \left( \phi^*_p - \Psi_{RP}\right)\rangle_{flow} = 
\int {\frac{d\Omega^*_p}{4\pi} \frac{d\phi_H}{2\pi}
A({\bf p}_H, {\bf p}^*_p) \sin \left( \phi^*_p - \phi_H\right)}
v_1^H ({\bf p}_H)~~.
\end{eqnarray}
This contribution is additive in the hyperons global polarization measurement.
At fixed hyperon pseudo-rapidity and transverse momentum,
this term is defined by the product of the hyperons directed flow and the following function:
\begin{eqnarray}
\label{AcorrV1}
A_{corr}^{flow}(p_t^H,\eta^H) =
\int {\frac{d\Omega^*_p}{4\pi} \frac{d\phi_H}{2\pi}
A({\bf p}_H, {\bf p}^*_p) \sin \left( \phi^*_p - \phi_H\right)}.
\end{eqnarray}
This integral could be non-zero due to detector acceptance%
\footnote{It could be also non-zero contribution to this integral from transverse polarization of the hyperon.
To distinguish this contribution from detector effect require more assumptions and
such consideration is out of scope of this paper.}.
Taking into account that directed flow is an anti-symmetric function of pseudorapidity,
for an estimate we will consider asymmetry of this function (\ref{AcorrV1}),
which we define as a difference between the integral over positive and negative hyperons pseudo-rapidity regions:
$A^{flow}_{asym}(p_t^H) = \left[\int_{\eta^H>0}-\int_{\eta^H<0}\right]A_{corr}^{flow}(p_t^H,\eta^H)$.
Fig.\ref{v1asymmetry} shows the asymmetry $A^{flow}_{asym}$
calculated for the $\Lambda$ hyperons reconstructed from Au+Au collisions at $\sqrt{s_{NN}}$=200 GeV with the STAR detector at RHIC.
Setting as an estimate the $\Lambda$ directed flow to be 10\%,
the flow contribution (\ref{meanSinV1}) appears to be less than $2\times 10^{-3}$.
This is an order of magnitude smaller that the upper limit for the 
$\Lambda$ global polarization obtained in \cite{Selyuzhenkov:2005xa,Selyuzhenkov:2006fc}
and it is negligible in this global polarization measurement.

\paragraph{\label{Conclusion}Conclusion}
We derived equations for acceptance correction function
for hyperons global polarization measurement.
For the case of the $\Lambda$ hyperons measured with the STAR detector at RHIC,
the deviation of this corrections from unity is found to be small
and they could change the absolute value of the measured $\Lambda$ hyperons  global polarization by less than 20\%.
We also investigated contribution from hyperon's directed flow
to the global polarization measurement and found it negligible in the measurement
with $\Lambda$ hyperons reconstructed with the STAR detector at RHIC.

\end{document}